\documentclass[sigconf,nonacm]{acmart}
\usepackage{tikz}
\AtBeginDocument{%
  \providecommand\BibTeX{{%
    \normalfont B\kern-0.5em{\scshape i\kern-0.25em b}\kern-0.8em\TeX}}}

\makeatletter
\def\@ACM@checkaffil{
    \if@ACM@instpresent\else
    \ClassWarningNoLine{\@classname}{No institution present for an affiliation}%
    \fi
    \if@ACM@citypresent\else
    \ClassWarningNoLine{\@classname}{No city present for an affiliation}%
    \fi
    \if@ACM@countrypresent\else
        \ClassWarningNoLine{\@classname}{No country present for an affiliation}%
    \fi
}
\makeatother

\begin{document}

\title{S3C2 Summit 2023-02: \\ Industry Secure Supply Chain Summit}

\author{Trevor Dunlap$^{*}$, Yasemin Acar$^{\dagger}$, Michel Cukier$^{\ddagger}$, William Enck$^{*}$,\\ Alexandros Kapravelos$^{*}$, Christian Kästner$^{\mathsection}$, Laurie Williams$^{*}$}

\def \authors{Mindy Tran, Yasemin Acar, Michel Cukier, William Enck, Alexandros Kapravelos, Christian Kästner, Laurie Williams}

\affiliation{%
    \institution{ $^*$North Carolina State University, Raleigh, NC, USA}
}
\affiliation{%
    \institution{$^\dagger$Paderborn University, Paderborn, Germany and George Washington University, DC, USA}
}
\affiliation{%
    \institution{$^\ddagger$University of Maryland, College Park, MD, USA}
}
\affiliation{%
    \institution{ $^\mathsection$Carnegie Mellon University, Pittsburgh, PA, USA}
}

\renewcommand{\shortauthors}{Secure Software Supply Chain Center (S3C2)}
\renewcommand{\shorttitle}{S3C2 Summit 2023-02: Industry Secure Supply Chain Summit}

\begin{abstract}
  Recent years have shown increased cyber attacks targeting less secure elements in the software supply chain and causing fatal damage to businesses and organizations. Past well-known examples of software supply chain attacks are the SolarWinds or log4j incidents that have affected thousands of customers and businesses. The US government and industry are equally interested in enhancing software supply chain security. On February 22, 2023, researchers from the NSF-supported Secure Software Supply Chain Center (S3C2) conducted a Secure Software Supply Chain Summit with a diverse set of 17 practitioners from 15 companies.  The goal of the Summit is to enable sharing between industry practitioners having practical experiences and challenges with software supply chain security and helping to form new collaborations.
  We conducted six-panel discussions based upon open-ended questions regarding 
  software bill of materials (SBOMs), 
  malicious commits, 
  choosing new dependencies,
  build and deploy, 
  the Executive Order 14028, and 
  vulnerable dependencies. 
  The open discussions enabled mutual sharing and shed light on common challenges that industry practitioners with practical experience face when securing their software supply chain. In this paper, we provide a summary of the Summit. Full panel questions can be found at the beginning of each section and in the appendix.
\end{abstract}




\maketitle

\begin{tikzpicture}[overlay, remember picture]
\node[anchor=north west, 
      xshift=17.5cm, 
      yshift=-2.1cm] 
     at (current page.north west) 
     {\includegraphics[width=2.1cm]{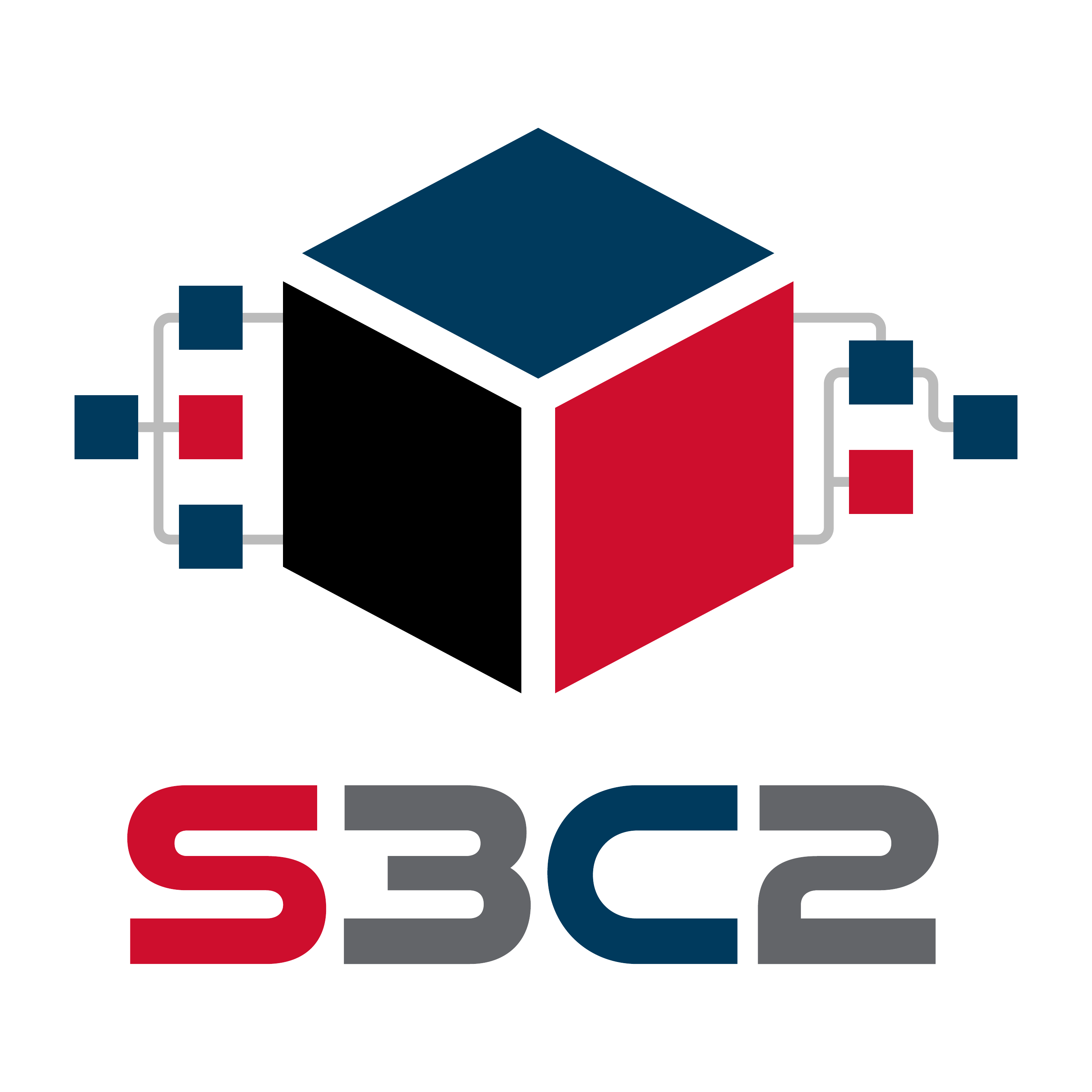}}; 
\end{tikzpicture}

\section{Introduction}

  Recent years have shown increased cyber attacks targeting less secure elements in the software supply chain and causing fatal damage to businesses and organizations. Past well-known examples of software supply chain attacks are the SolarWinds or log4j incidents that have affected thousands of customers and businesses. On February 22, 2023, four NSF-supported Secure Software Supply Chain Center (S3C2)\footnote{\url{https://s3c2.org}} researchers conducted a one-day Secure Software Supply Chain Summit with a diverse set of 17 practitioners from 15 companies.  The goal of the Summit is to enable sharing between industry practitioners having practical experiences and challenges with software supply chain security; to help form new collaborations between industrial organizations and researchers; and to identify research opportunities.

  Summit participants were recruited from 15 companies, intentionally in diverse domains and having various company maturity levels and sizes \textemdash ten major corporations, four medium size companies, and one start-up).  Except for the host company with three participants, all companies could only have one participant.  Attendance is limited to one per company to keep the event small enough that honest communication between participants can flow. The Summit was conducted under the Chatham House Rules, which state that all participants are free to use the information discussed, but neither the identity nor the affiliation of the speaker(s), nor any other participant may be revealed. As such, none of the participating companies are identified in this paper.  

  The Summit consisted of one keynote presentation and six panels. Before the Summit, participants completed a survey to vote on the topics of the six panels.  As such, the panel topics represent the challenges faced by practitioners. Based upon personal preferences expressed in the survey, four participants were selected to begin each 45-minute panel discussion with a 3-5 minute statement.  The remaining minutes of each panel were spent openly discussing the topic.  The questions posed to the panelists appear in the Appendix.   

  The four researchers (three professors, one Ph.D. student), and several participants took notes on the discussions.  The Ph.D. student created a first draft summary of the discussion based on these notes.  The draft was first reviewed by the three professors at the Summit and then by the three other authors of this paper, who are also S3C2 researchers and experts in software supply chain security.

  The next seven sections provide a summary of the Secure Supply Chain Summit.

\section{Executive Order}
Executive Order 14028 \cite{EO} issued on May 12, 2021, charges organizations supplying critical software to the US government to improve the security and integrity of their software and the software supply chain.  Most organizations need to make procedural, operational, and cultural changes. 



\subsection{More Work To Do}

The participants had a range of reactions to the EO, particularly because not all of the participant's organizations sell software to the US government.
The first panelist felt that the EO has been beneficial because it helped get leadership buy-in for adopting security practices they have wanted to adopt for a while.
The EO is helping to create structure and efforts that span company divisions, including sales.
The second panelist had a different experience.
They wished they could say that the EO translated into additional resources for them.
However, they have seen pushback from leadership in providing software bill of materials (SBOMs).
Their customers do not get SBOMs directly.
SBOMs are given to governments when required, and the process goes through their company's cybersecurity office.
Another participant noted that they are not feeling the same urgency as they experience with complying with General Data Protection Regulation (GDPR).
There were differing opinions on what the urgency is and when the impact will be felt.
Some felt the urgency is being felt now, while others did not anticipate feeling urgency for two to three years.

Several participants raised concerns about the vagueness of the EO and the challenges they encountered in trying to comply with it.
Questions arose surrounding the scope of applicability and the feasibility of achieving a ``zero vulnerability'' policy.
One participant noted that buying something vulnerability-free is impossible because new vulnerabilities are reported daily.

Finally, one participant familiar with drafts of the upcoming National Cybersecurity Strategy was more positive.
They believe there will be a positive upheaval in how we think about cyber insurance and liability, noting that, ``Sunlight is the best disinfectant -- the fact that you have to get it out there in the open will drive better behavior.''
They were optimistic that the strategy change would finally break the log jam.

\subsection{Open Questions}
At the end of the panel, some open questions remained:

\begin{itemize}
\item What are the attestation requirements for the executive order?
\item How do we move the executive order beyond a compliance checkbox?
\end{itemize}

\section{Software Bill of Materials (SBOM)}
An SBOM is a nested inventory of ``ingredients'' that make up the software component or product that helps to identify and keep track of third-party components of a software system. The EO \cite{EO} states that any company that sells software to the federal government is mandated to issue a complete SBOM that complies with the National Telecommunications and Information Administration (NTIA) Minimal Elements \cite{NtiA}.  


\subsection{Concerns about Utility}

The four panelists were generally positive about the SBOM effort, but most felt the current state of SBOM was more of a fairly meaningless ``compliance-check-the-box'' providing a bunch of stats and information that no one will look at.
One panelist suggested that the real value of SBOMs is that they give vendors embarrassment-motivation to not put out an SBOM ``with bad stuff in it.''
This panelist suggested that the executive order will likely be considered a false start for the software supply chain.

A participant recently did an audit of available SBOMs.
They noted that nearly none met the NTIA minimum requirements \cite{NtiA}.
In some cases, fields are hard to fill, e.g., the ``supplier'' for open-source software.
The notion of a ``minimum SBOM'' was also questioned, with one participant indicating that a minimum SBOM does not need to include transitive dependencies.
They did not think such an SBOM was useful.

One participant noted that they were struggling to use the SBOM for incident response within the company and were unsure how to tell a customer to use it.
Another participant noted that reproducing SBOMs is hard.
Another noted that they have seen SBOMs that are gigabytes in size and often see lots of fields that say ``no assertions.''
Others noted concerns that developers would likely game SBOMs by simply renaming packages or folder names.
There was a feeling that SBOM has become a catchword and has caught on too much.
The industry is missing that the point is to address vulnerabilities.
One participant noted that if we put more focus on addressing vulnerabilities, then in 15 years, we may not need an SBOM.

\subsection{Vulnerability Exploitability eXchange (VEX)}

Vulnerability Exploitability eXchange (VEX) \cite{VEX} information was touched upon throughout the discussion.
There was a consensus that the manual nature of VEX information will significantly limit its value.
For example, one participant noted that it would be very difficult to use in automated processes.
As such, they questioned why we are so focused on providing it.
Another participant noted that VEX information may be another indicator of code quality.
The participant suggested that if an SBOM discloses 10,000 ``I'm not exploitables,'' a code quality issue may be suspected.

\subsection{Releasing to Public}

The participants had mixed opinions about releasing SBOMs to the public.
Some were concerned that SBOMs provide attackers significant lead time by providing a blueprint down to the version level.
Others were less concerned.
One participant noted that Jupiter Networks had put their SBOM online for years.
Does this put them at more risk?
Possibly, but a different kind of risk.
In general, transparency has more benefits than no transparency.

Later discussion also noted that there are different levels of generating an SBOM: source SBOM, build SBOM, or deploy SBOM.
It was unclear where we should draw the line in what we give, whether to customers or the public.

\subsection{Fixing Vulnerabilities}

Participants discussed challenges around fixing vulnerabilities.
There was discussion around requirements that companies do not ship with known vulnerabilities.
A participant stated that the vulnerabilities that need to be fixed are  \emph{publicly}-known vulnerabilities, noting that it takes time to fix the items that are known internally.
Another participant noted that everything comes back to culture:
Developers are technical artists, and you cannot tell them they need to use specific tools or mandate how you notify them about vulnerabilities.
We have to identify technical debt that has an impact versus technical dept that affects the artistry.

\subsection{Open Questions}
At the end of the panel, some open questions remained:

\begin{itemize}
\item How can SBOMs handle when vendors fork and rename the dependencies? Defeating the entire toolchain and avoiding updates.
\item How can we make the VEX process (production and consumption) less manual?
\item What are typical remediation practices and timelines for handling vulnerabilities identified within SBOMs?
\item How do we identify what is worth fixing and what isn’t worth fixing?
\end{itemize}

\section{Choosing Dependencies}

Open source dependencies vary widely in quality, maintenance, origin, and licenses.
Every dependency introduces value and risk, and 
once it is incorporated into a project, it is often hard to replace.
Therefore, it is important to have a policy that governs how software developers may choose new dependencies.


\subsection{No Easy Metrics}

Participants discussed a range of strategies for choosing dependencies, and it was apparent that there were no good metrics.
One participant noted that they gravitate towards more widely-used dependencies.
They have checklists about using dependencies, originally from a legal perspective, but it is unclear how robust the checklists are.
Another participant noted that there is a need to support the artistry of the work.
They have a lightweight process that is focused on quick turnaround.
When something is not approved, they engage with developers to help find an alternative.

Other discussed topics included known CVEs, the number of dependencies, company backing, and popularity.
However, there was caution in making decisions on those aspects alone.
For example, having no CVEs may be a red flag, whereas having many CVEs also might be a good sign.
For example, lots of CVEs often occur after a project begins to run security tools against their code regularly.
One participant noted giving weight to dependencies that have zero external dependencies.
Another participant cautioned that this might be a sign that the project simply copied external code (e.g., crypto) into their project, which could cause problems later.
Company backing of a dependency is also not always seen as helpful.
A participant noted that some large companies open source thousands of projects and then forget about them.
Even popular projects have their downsides.
Several participants raised concerns about projects with a ``benevolent dictator for life'' and the challenges of incorporating security changes.
Ultimately, the decision to choose a dependency requires a qualitative feel.

The panel discussion also touched upon the use of OpenSSF Scorecard \footnote{https://github.com/ossf/scorecard} security health metrics.
Of the represented companies, only one or two used Scorecard as a metric.
However, many others looked at it and evaluated how they might incorporate it.
One participant noted competing scorecards that seem different (e.g., from Synk and Coverity).
There was a sentiment of encouragement for more participation in OpenSSF working groups.

The keynote speaker discussed picking better components, where ``better'' means ``less likely to have a vulnerability.''
The speaker said that they looked at OpenSSF Scorecard scores and measured whether or not packages were less likely to have vulnerabilities if their score was higher.
Unfortunately, they found no connection.
They then used machine learning to find trends in the different OpenSSF Scorecard categories.
The most predictive practice was code review, followed by not checking in binaries and then pinning dependencies.
The least important factor was using fuzzing.

\subsection{Investing back in the community}

Organizations using open-source software need to realize they are investing.
One participant noted:
(1)~you get what you pay for;
(2)~people value things at the price they pay for them; and
(3)~community open source is free if your time is worthless (you are paying your developers to use it).
The discussion continued, arguing for a difference between products and projects: 
Products are backed and sold, whereas projects are community-based. 
If the industry wants to use open-source projects within their products, they must realize the responsibility is back on them to ensure the project is updated.

How this contribution occurs is challenging.
One participant noted that even medium-sized companies do not have resources for dedicated people to work on helping open-source projects.
Contributions generally happen out of passion from the developers outside their normal roles.

\subsection{Open Questions}

\begin{itemize}
    \item Can the current qualitative feel used to choose dependencies be turned into a quantitative metric? (and should it?)
    \item How can companies better support the open source community to avoid a tragedy of the commons?
\end{itemize}

\section{Detecting Malicious Commits}
Actors of past software supply chain attacks (e.g., SushiSwap) use malicious commits to submit unauthorized changes to the source repository. Detecting and discerning these malicious commits is not always straightforward. Attackers often use obfuscated code, steal authentication credentials, or use various impersonation strategies to deceive and put malicious code changes through the system. 


\subsection{Intent}

Overall, the panelists and participants did not have good solutions for detecting malicious commits.
They mentioned several standard mitigations, including two-person reviews, a list of maintainers who can merge into the main branch, and a set of trusted gatekeepers.
There was also a hope that linters and scanners could catch some low-hanging fruit, such as malicious binaries checked into repositories.
However, all of these approaches have limitations.
One participant noted that even a two-person review is limited, noting that they have seen questionable commits between three developers who sit close to each other.
Participants also discussed the difficulty of identifying intent.
Is code poorly written or malicious?
In some cases, such as protestware and typosquatting, the intent is clearly malicious, but in many other cases, it is not.

Participants also discussed the impact of advances in artificial intelligence (AI) and machine learning (ML).
They noted that Copilot is making developers significantly more productive.
There is already evidence that the AI programming assistant Copilot\footnote{https://github.com/features/copilot} can suggest vulnerable code because it has been trained on vulnerable code.
There was a worry about such tools being either intentionally or unintentionally trained on malicious code.
On the flip side, ML was also seen as a solution.
There was hope that there would be enough signal to use ML for outlier detection. 
For example, asking ``how often does this developer commit to this part of the code base?'' Participants essentially asked for a scorecard approach for committers. 

\subsection{Open Questions}
At the end of the panel, some open questions remained:

\begin{itemize}
\item What monitoring techniques should we use to detect if others are updating code that should not be updated? 
\end{itemize}

\section{Secure Build and Deploy}
Various build platforms and CI/CD tools support developers in automating the parts of software development related to building, testing, and deploying. These platforms further help in enhancing software build integrity by establishing documented and consistent build environments, isolating build processes, and generating verifiable provenance. 


\subsection{Making the Transition}

Several panelists have invested significant time and resources in securing their build and deploy processes.
Supply-chain Levels for Software Artifacts (SLSA)~\cite{slsa} was promoted as a great framework.\footnote{Note that SLSA v0.1 was the latest version of SLSA at the time of the summit.}
A representative from a relatively new company said that they were SLSA level ``3.7''.
They started with reproducibility and saw it as essential to get right from the beginning, noting that reproducible builds are hard to bolt on afterward.
Interestingly, this participant saw hermetic builds environments as more challenging than reproducible builds.
A hermetic build is when the build is accomplished without a network connection. 

A representative of an older company echoed the challenges of dealing with legacy build environments.
This participant noted that their company has two pipelines: one for the old projects and one for the new projects.
The new pipeline focuses on security: it removes service accounts and limits how dependencies can be added.
There has been a great effort to remove the human element as much as possible.
For example, developers are not allowed to write Dockerfiles.
Instead, they use templates that generate Dockerfiles, which are in a versioning system to prevent developers from sneaking in changes.
As more projects are on-boarded to the new pipeline, they start learning about more and more dependencies.
This results in project discussions about choosing dependencies, and if approved, new dependencies are brought into Artifactory to make them available to a hermetic build pipeline.

The discussion also highlighted the importance and challenges of other aspects of SLSA.
The importance of ephemerality was highlighted many times, with one participant noting that it can help to identify and remove long-lived secrets.
However, key management is challenging.
Some participants noted using short-lived JWTs and zero trust in their pipeline. 
A participant noted using Sigstore as another option to guarantee certain aspects have been signed rather than managing keys at the start of a build project.
The discussion also touched on the value of Hardware Security Modules (HSMs) for code signing keys and cloud-based substitutes.

Ephemerality is particularly tricky when incorporating tests.
One participant noted that for products with multiple network-connected components, it is impossible to have ephemeral versions of those components available in the CI pipeline.
Another participant noted that their projects often interact with specialized hardware, which prevents them from achieving full ephemerality.

\subsection{Reproducible Builds}

In contrast to the 2022-09 summit~\cite{Summit1}, the summit participants exhibited excitement around the topic of reproducible builds.
As noted above, one panelist said they achieved reproducible builds before hermetic builds.
The company with multiple pipelines is not yet achieving reproducible builds but is interested in getting there because it provides even more determinism in the build process.
Participants noted that Dockerfiles are particularly problematic for reproducible builds and suggested using alternatives if language dependencies support them.
For example, ko\footnote{\url{https://github.com/ko-build/ko}} is a simple container image for Go applications.

\subsection{Open Questions}
At the end of the panel, some open questions remained:

\begin{itemize}
\item How to best handle key rotation for cloud-based builds?
\item What properties are we looking to get out of hermetic builds?
\item What are ways to automatically identify why pipelines generate different builds? (i.e., nonreproducible builds)
\item How can we better educate computer science students about securing builds?
\end{itemize}

\section{Updating Vulnerable Dependencies}

Most software uses a plethora of third-party dependencies that provide common functionality and help developers with productivity. However, these dependencies add complexity and lead to a vast ecosystem of (transitive) dependencies that each software replies on. A security vulnerability in a third-party dependency can lead to cascading issues and needs to be updated with the newly released version fix as soon as possible. Companies commonly rely on different strategies and tools when updating vulnerable dependencies. 


\subsection{Knowing Your Dependencies}

Software Composition Analysis (SCA) tools used to scan projects for vulnerabilities are widely used and increasingly part of the company culture.
One participant said they use their SCA tool for every git push and pull request; it is a key component of their overall ML-based risk profile tools.
A second panelist had a more pessimistic view, noting that finding vulnerabilities is a net negative value in that ``liabilities'' are created because we cannot ``unsee'' them.
The Equifax employees did not get fired because they got hacked \textemdash
they got fired because they knew about the problem six months before they got hacked.

The keynote speaker reported that SCA tools reduce risk by 22\%, which is lower than hoped for.
Unfortunately, just telling developers where vulnerabilities are does not mean it improves their ability to stay up-to-date and manage patches efficiently.
The speaker noted that media coverage often impacts the response time for addressing critical vulnerabilities.
Ultimately, the math on dependency management is daunting.
The average Java project has 150 dependencies.
Each dependency has, on average, 10 releases per year, resulting in 1,500 updates to consider.
Any of these updates might introduce a breaking change.
Organizations need a process for staying up to date.

The second panelist suggested that having a good automated test suite is the real challenge.
While there is a growing suite of tools,
the real value is the rapid resolution of vulnerabilities.
But how can you be confident when updating a dependency? 
The common technique is time-consuming: scoping, scheduling, and manually correcting. 
The community needs to put more effort behind automated functional test suites because, without that, organizations cannot have the confidence to rapidly update to the latest versions. 

Three main techniques exist for facing vulnerable dependencies: 
you remove, patch, or upgrade the dependency. 
Participants noted that the preferred option is to remove the dependency if it is unnecessary. 
The other two are much more realistic if it is a true dependency. Vulnerability management becomes easier when you only include what you need. 
Participants noted that debloating containers at the system layer reduces the attack surface, increases boot-up time, and lowers cloud service bills.
SCA tools are not only helpful in discovering vulnerabilities but also in what you do not use.
Participants discussed a lot of techniques for reducing container images, after which one participant noted, 
``We spent a lot of time today talking about container issues.
It's at the open source library dependency level where we don't have great answers.''

\subsection{Open Questions}
At the end of the panel, some open questions remained:

\begin{itemize}
\item How can we further drive towards an automated functional test suite?
\item How can we efficiently eliminate unneeded library dependencies?
\end{itemize}

\section{Current state of the Software Supply chain}
\label{misc}

The keynote speaker discussed the current state of the software supply chain.
The presentation began with a discussion of malicious packages in the supply chain, focusing on npm, which is used 3-4x more than other package ecosystems.
The speaker observed that malicious attackers are taking over maintenance, gaining the trust of development teams, and using typosquatting to corrupt the open-source ecosystem.
The speaker cited Sonotype's report \cite{sonatype_2022} of a 742\% growth year over year, with over 97K malicious packages discovered as of October 2022.
These malicious packages generally fall into four categories: dependency confusion, typosquatting, protestware, and malicious code injections, with dependency confusion and typosquatting being the two main drivers.
It makes sense for these categories to dominate the numbers, as their success primarily stems from injecting many malicious packages.
One interesting takeaway from the discussion is that current attacks have fairly low sophistication, suggesting that our current defenses similarly lack sophistication.
For example, many typosquatting attacks are ``smash and grab'' attempts, where the attacker is looking to execute code on the developer's host (e.g., via pre-install scripts). The code in the package itself is often trivial or broken.

The keynote speaker then discussed perceptions versus reality in open-source risk management.
This discussion was primarily based on Java and Maven Central.
The speaker noted a general perception that open source is risky.
For example, in Maven Central, 35\% of releases are vulnerable, equating to 3.4M vulnerable releases and 1.2B vulnerable downloads per month.
However, in reality, open-source software can almost always be secure: 98\% of projects have versions without known vulnerabilities available, and most vulnerabilities are patched before they are disclosed.
Vulnerabilities are largely a consumption-side problem: for 96\% of vulnerable downloads, a non-vulnerable version was available.
Getting developers to stay up to date and maintain their dependencies will significantly decrease risks.
However, this maintenance may not be practical or realistic.

The second presented perception is that the industry is good at managing open source.
Surveys suggest that 68\% of companies are confident that they are not using vulnerable versions of software and self-report high remediation maturity.
However, in reality, 68\% of applications use a component with a known vulnerability.
Managers like to say that remediation is fast, but it took 52 days to reach 70\% remediation of log4shell.
One participant hypothesized that the delay results from a project mentality around software rather than a product mentality: once a project is done and shipped, the team is on to making new things.
Another participant suggested that developers often have the attitude that once no one is working on a piece of code anymore, it does not matter if the code it uses is deprecated.
The discussion also turned to the difference between language ecosystems.
Java is inflexible, requiring developers to manually update versions, whereas Node is too flexible.
Newer languages, such as Go, are learning from the past and striking a balance.
\section{Executive Summary}
\label{executive}
Some participants who sell software to the government have found the EO \cite{EO} useful for increasing the security focus of their company. Other participants have not experienced a similar urgency while all wanted to prevent the EO from becoming a compliance checkbox.  Participants were generally positive about producing SBOMs though concerns were expressed that current SBOM generation tools do not meet the NTIA \cite{NtiA} minimum requirements and about storing large SBOM files and the risks of sharing SBOMs, such as providing attackers information. Participants discussed strategies for choosing dependencies with lower security risk, including using quantitative metrics, such as the OpenSSF Scorecard metrics.  The discussion indicated that better strategies for choosing dependencies need to be developed and that companies should consider how to give back to the open-source community.   

Attackers are increasingly injecting malicious components into ecosystems. Like participants of the 2022-09 summit~\cite{Summit1}, participants acknowledged that detecting contributions with malicious intent is difficult.  They were hopeful that AI/ML might be used to detect malicious contributions but also were concerned about code generation tools being trained on vulnerable code.  Participants found value in the SLSA~\cite{slsa} practices for securing the build environment, isolating build processes, and generating verifiable provenance.  Participants indicated that securing the build process of legacy products was more difficult than establishing a more secure pipeline for newer products. Compared with the participants of the 2022-09 summit~\cite{Summit1}, the participants in this summit were more positive about reproducible builds.  Participants discussed the impracticalities of updating vulnerable direct and transitive dependencies, even assisted by SCA tools.  Solutions included increasing automated testing suites to give more confidence that a new version of a dependency does not break a system and by removing dependencies that are not used.

\begin{acks}
A big thank you to all Summit participants. We are very grateful for being able to hear about your valuable experiences and suggestions. The Summit was organized by Laurie Williams, William Enck, and Yasemin Acar and was recorded by Trevor Dunlap.  This material is based upon work supported by the National Science Foundation Grant Nos. 2207008, 2206859, 2206865, and 2206921.
These grants support the Secure Software Supply Chain Summit (S3C2), consisting of researchers at North Carolina State University, Carnegie Mellon University, University of Maryland, George Washington University, and Paderborn University. Any opinions expressed in this material are those of the author(s) and do not necessarily reflect the views of the National Science Foundation.
\end{acks}

\bibliographystyle{ACM-Reference-Format}
\bibliography{literature}


\begin{thebibliography}{6}


\ifx \showCODEN    \undefined \def \showCODEN     #1{\unskip}     \fi
\ifx \showDOI      \undefined \def \showDOI       #1{#1}\fi
\ifx \showISBNx    \undefined \def \showISBNx     #1{\unskip}     \fi
\ifx \showISBNxiii \undefined \def \showISBNxiii  #1{\unskip}     \fi
\ifx \showISSN     \undefined \def \showISSN      #1{\unskip}     \fi
\ifx \showLCCN     \undefined \def \showLCCN      #1{\unskip}     \fi
\ifx \shownote     \undefined \def \shownote      #1{#1}          \fi
\ifx \showarticletitle \undefined \def \showarticletitle #1{#1}   \fi
\ifx \showURL      \undefined \def \showURL       {\relax}        \fi
\providecommand\bibfield[2]{#2}
\providecommand\bibinfo[2]{#2}
\providecommand\natexlab[1]{#1}
\providecommand\showeprint[2][]{arXiv:#2}

\bibitem[CISA(2022)]%
        {VEX}
\bibfield{author}{\bibinfo{person}{CISA}.} \bibinfo{year}{2022}\natexlab{}.
\newblock \showarticletitle{Vulnerability Exploitability eXchange (VEX)}.
\newblock
  \bibinfo{journal}{\emph{https://www.cisa.gov/sites/default/files/publications/VEX\_Use\_Cases\_Document\_508c.pdf}}
  (\bibinfo{year}{2022}).
\newblock


\bibitem[House(2021)]%
        {EO}
\bibfield{author}{\bibinfo{person}{US~White House}.} \bibinfo{year}{May 12,
  2021}\natexlab{}.
\newblock \showarticletitle{Executive Order 14028 on Improving the Nation's
  Cybersecurity}.
\newblock
  \bibinfo{journal}{\emph{https://www.whitehouse.gov/briefing-room/presidential-actions/2021/05/12/executive-order-on-improving-the-nations-cybersecurity/}}
  (\bibinfo{year}{May 12, 2021}).
\newblock


\bibitem[NTIA(2021)]%
        {NtiA}
\bibfield{author}{\bibinfo{person}{NTIA}.} \bibinfo{year}{July 21,
  2021}\natexlab{}.
\newblock \showarticletitle{The Minimal Elements of a Software Bi,ll of
  Materials}.
\newblock
  \bibinfo{journal}{\emph{https://www.ntia.doc.gov\/files\/ntia\/publications\/sbom\_minimum\_elements\_report.pdf}}
  (\bibinfo{year}{July 21, 2021}).
\newblock


\bibitem[OpenSSF(2023)]%
        {slsa}
\bibfield{author}{\bibinfo{person}{OpenSSF}.} \bibinfo{year}{2023}\natexlab{}.
\newblock \showarticletitle{Supply-chain Levels for Software Artifacts (SLSA)}.
\newblock \bibinfo{journal}{\emph{https://slsa.dev/}} (\bibinfo{year}{2023}).
\newblock


\bibitem[Sonatype(2022)]%
        {sonatype_2022}
\bibfield{author}{\bibinfo{person}{Sonatype}.} \bibinfo{year}{2022}\natexlab{}.
\newblock \showarticletitle{700\% Average Increase in Open Source Supply Chain
  Attacks}.
\newblock
  \bibinfo{journal}{\emph{https://www.sonatype.com/press-releases/sonatype-finds-700-average-increase-in-open-source-supply-chain-attacks}}
  (\bibinfo{year}{2022}).
\newblock


\bibitem[Tran et~al\mbox{.}(2022)]%
        {Summit1}
\bibfield{author}{\bibinfo{person}{Mindy Tran}, \bibinfo{person}{Yasemin Acar},
  \bibinfo{person}{Michel Cucker}, \bibinfo{person}{William Enck},
  \bibinfo{person}{Alexandros Kapravelos}, \bibinfo{person}{Christian Kastner},
  {and} \bibinfo{person}{Laurie Williams}.} \bibinfo{year}{Sept
  2022}\natexlab{}.
\newblock \showarticletitle{S3C2 Summit 2202-09: Industry Secure Suppy Chain
  Summit}.
\newblock \bibinfo{journal}{\emph{http://arxiv.org/abs/2307.15642}}
  (\bibinfo{year}{Sept 2022}).
\newblock


\end{thebibliography}

\appendix

\section{Full Survey Questions for Panel}
\label{questions}
\begin{enumerate}
\item Where are you in your journey toward producing an SBOM?  What will/can SBOMs actually achieve? Are they a waste of time?  How can they be leveraged/used?  Are you creating a VEX?  How?
\item How can malicious commits be detected? What do you think signals a suspicious/malicious commit?  What role does the ecosystem play in detecting malicious commits?
\item Are you more careful now in choosing new dependencies?  Do you use OpenSSF Scorecard or other metrics to help you make decisions? 
\item What is being done (or should be done) to secure the build and deploy process/tooling pipeline?  Are you working toward reproducible builds?
\item What changes is your company making in relation to the Executive Order?
\item What is your process for updating vulnerable dependencies?   Do you always keep up-to-date?  What kind of testing or other strategies do you use before updating to a new version?  How is the output of an SCA tool used?
\end{enumerate}

\end{document}